\documentclass{sf2a-conf2012}
\usepackage{graphicx}
\usepackage{hyperref}
\usepackage[]{natbib}  
\usepackage[cyr]{aeguill}
\usepackage{epstopdf}

\def\BibTeX{{\rm B\kern-.05em{\sc i\kern-.025em b}\kern-.08em
    T\kern-.1667em\lower.7ex\hbox{E}\kern-.125emX}}
\bibpunct{(}{)}{;}{a}{}{,}  %%%%%%%%%%%%%  A&A bibliography style
\begin{document}

\TitreGlobal{SF2A 2012}

%%-----------------------------------------------------------------
%%      the top matter
%%

\title{Constraining the nature of the Galactic center black hole Sgr~A* with present and future observations}

\runningtitle{Constrinaing the nature of Sgr~A*}

\author{F. H. Vincent}\address{Laboratoire AstroParticule et Cosmologie, CNRS, Universit\'e Paris Diderot, 10 rue Alice Domon et L\'eonie Duquet, 75205 Paris Cedex 13, France}

\author{E. Gourgoulhon}\address{LUTH, Observatoire de Paris, CNRS, Universit«e Paris Diderot, 5 place Jules Janssen, 92190 Meudon, France}

\author{O. Straub}\address{Nicolaus Copernicus Astronomical Center, Bartycka 18, 00-716 Warsaw, Poland}

\author{M. Abramowicz}\address{Department of Physics, G\"oteborg University, SE-412-96 G\"oteborg, Sweden}

\author{J. Novak$^{2}$}

\author{T. Paumard}\address{LESIA, Observatoire de Paris, CNRS, Universit«e Paris Diderot, 5 place Jules Janssen, 92190 Meudon, France}

\author{G. Perrin$^{5}$}

%% IF Author3 has the same affiliation than Author1:
%\author{C.\,E. Author3$^1$}

%% IF Author3 has its own affiliation:
%\author{C.\,E. Author3}\address{Dept. of Chess, University of Games, 35101 Las Vegas, Monaco} 

%% IF Author3 has two affiliations, the one of Author1 and a second one:
%\author{C.\,E. Author3$^{1,}$}\address{Dept. of Chess, University of Games, 35101 Las Vegas, Monaco} 

%% Keep this line, even if the page will be settled afterwards.
\setcounter{page}{237}

%%-----------------------------------------------------------------

\maketitle

%%-----------------------------------------------------------------
%%        The abstract
%% 
%%  Warning!  within the abstract:
%%  - do not use macros. 
%%  - do not use commands like: \cite, \citet, \citep ... etc.

\begin{abstract}
The Galactic center is an ideal laboratory to study strong-field general relativistic phenomena, as the supermassive black hole Sgr~A* has the biggest angular Schwarzschild radius among all black holes. This article presents three different ways of using the immediate surroundings of Sgr~A* as a probe in order to either constrain its spin, or even test the very nature of this compact object. 
\end{abstract}

%% Insert the keywords (to appear in the ADS indexing)
%% Keywords must be separated by a comma

%%-----------------------------------------------------------------

\section{Introduction}
%%---------------------

The Galactic center is by far the closest galactic nucleus. It can thus be studied with great accuracy and give insights on the physics of galactic nuclei in general~\citep{genzel10}. Nearly four decades of observing campaigns have now made it extremely likely that the center of our Galaxy harbours a supermassive black hole, Sgr~A*~\citep{ghez08,gillessen09}. With its mass of $4.3\,\times10^{6}\,M_{\odot}$ at a distance of $8.3$~kpc from Earth, Sgr~A* has the biggest projected angular Schwarzschild radius among all black holes: $10\,\mu$as. This translates to an angular radius of the black hole shadow (i.e., to an apparent size of the event horizon) of around $24~\mu$as and an apparent angular radius of the ISCO of around $29~\mu$as for a spin of $0.5\,M$ ($M$ being the black hole mass)\footnote{Let us insist on the vocabulary used here. The \textit{projected} Schwarzschild radius is the angular size of the Schwarzschild radius that would be measured by a distant observer in a flat spacetime. The \textit{apparent} Schwarzschild radius is the actual angular size measured by a distant observer in the physical, curved spacetime. Due to general relativistic bending effects on the photons null geodesics, the apparent size is bigger than the projected size. For instance, in the Schwarzschild metric, the \textit{projected} radius of Sgr~A* event horizon is $10~\mu$as, whereas the \textit{apparent} radius of Sgr~A* event horizon (i.e., the radius of the shadow) is $3\sqrt{3}/2 \times 10 \approx 26~\mu$as. It is the \textit{apparent} size of an object, of course, that must be compared to the angular resolution of an instrument.} . Its surroundings are thus an ideal laboratory to study physics in extremely strong gravitational fields. Investigating strong-field general relativity (i.e. the effect of gravitation in the vicinity of compact objects) is one of the avenues for future tests of gravitation~\citep{will09}. Sgr~A* certainly is among the best targets to try performing such tests.

The aim of this paper is to discuss three different ways of constraining the nature of the compact object Sgr~A*, either with current or near-future observations: 

\begin{itemize}
\item[-] Fitting the spectrum emitted by the accretion structure surrounding Sgr~A* (section~\ref{torus}), 
\item[-] Constraining the black hole silhouette with future VLBI data (section~\ref{torus}),
\item[-] Simulating observations in the vicinity of alternative compact objects (section~\ref{3+1}).
\end{itemize}

A fourth way must be cited here: the near-future GRAVITY instrument~\citep{eisenhauer08,eisenhauer11} that will allow constraining Sgr~A*'s parameters by studying the dynamics of the Galactic center flares of radiation~\citep[see e.g.][and references therein]{vincent11d}. This fourth way will not be investigated here.

 \section{An ion torus surrounding Sgr~A*}
 \label{torus}
 
 The accretion structure that surrounds Sgr~A* has been shown to be part of the advection dominated accretion flows (ADAFs) as demonstrated by~\citet{narayan95a}. In this paper, this accretion structure is modeled by an ion torus~\citep{rees82}, derived from the Polish doughnuts class~\citep{abramowicz78}, the emitted radiation following the ADAF requirements given in~\citet{narayan95b}. This ion torus model is completely analytical and depends only on a few parameters with clear physical meaning: the black hole spin, the inclination parameter, the angular momentum value (assumed constant), the central energy and temperature, the ion to electron central temperature ratio and the magnetic to gas pressure ratio. The aim of this model is not to give a realistic description of the actual accretion structure surrounding Sgr~A*, that could only be described in details with GRMHD simulations. Instead, the ion torus model only aims at catching the main observable characteristics of the Galactic center accretion structure~\citep[see][for comparison of Polish doughnuts predictions and GRMHD simulations]{qian09}.
 
 A detailed description of the physics of the ion torus model can be found in~\citet{straub12}. Here, we only investigate the prospects of constraining the black hole parameters assuming its accretion structure can be correctly described by an ion torus. To do so, the emitted spectrum and the image of the torus are computed by means of the open source ray-tracing algorithm GYOTO\footnote{Available at \url{http://gyoto.obspm.fr}.}~\citep{vincent11a}.
 
\begin{figure}[ht!]
 \centering
 \includegraphics[width=0.48\textwidth,clip]{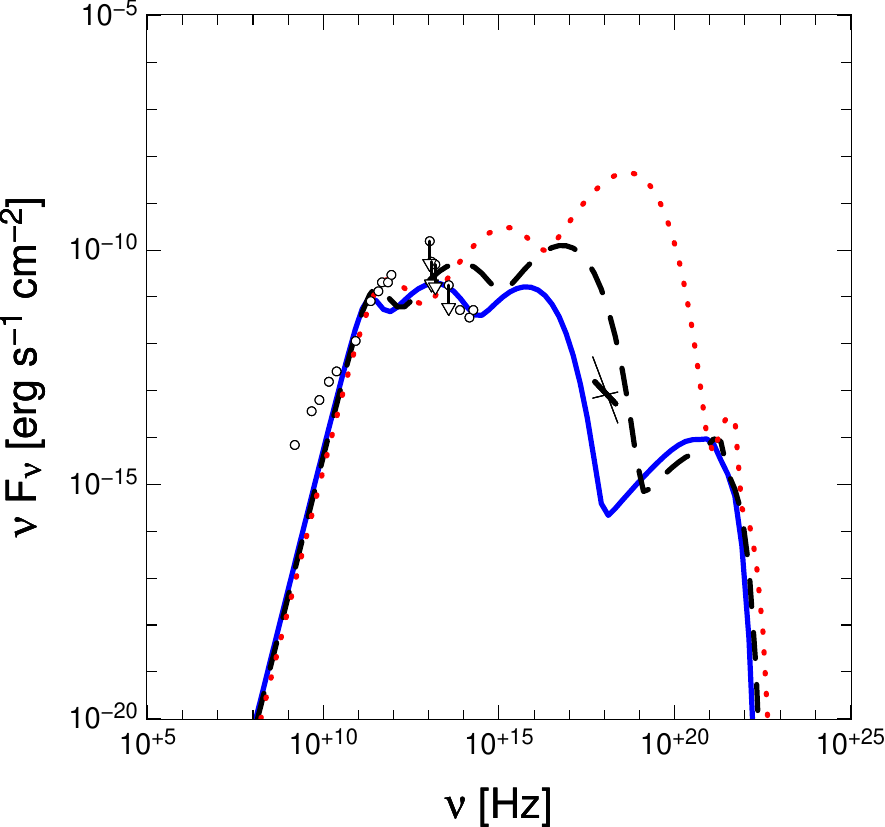}%      
 \includegraphics[width=0.48\textwidth,clip]{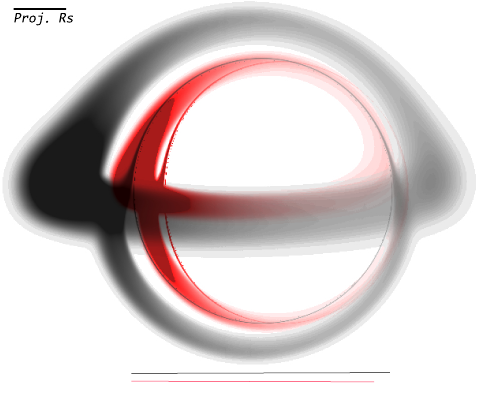}      
 \caption{Effect of the black hole spin parameter on the observables. {\bf Left:} Emitted spectrum of the ion torus. The spin parameter is $a = 0$ (solid blue), $0.5M$ (dashed black) or $0.9M$ (dotted red). References for the observed data points in black can be found in~\citet{straub12}. {\bf Right:} Superimposed images of the ion torus for a spin parameter $a=0.5M$ (black) and $0.9M$ (red). The angular size of the projected Schwarzschild radius ($10\,\mu$as) is given by the upper left solid line. The solid lines at the bottom of the figure show the angular diameters of the black hole silhouettes in both cases. The two angular diameters differ by an amount of approximately $3\,\mu$as. Adapted from~\citet{straub12}.}
  \label{fig:torus}
\end{figure}

 Fig.~\ref{fig:torus} shows the torus emitted spectrum and image for different values of the black hole spin, the other parameters being kept fixed at standard values, typical of the Galactic center accretion flow~\citep[see][for details]{straub12}. 
 
It appears that the ion torus is able of reproducing the order of magnitude of the very faint emission at the Galactic center~\citep[except at radio wavelengths, however see the discussion in][]{straub12}. The emitted spectrum is very similar at different spins for radio and infrared wavelengths. However, the X-ray part of the spectrum is highly dependent on the spin value. This is due to the fact that at higher spin, the torus shrinks and gets closer to the black hole: the flux is then shifted to higher energies. This dependency of the spectrum on the spin value makes it possible that Sgr~A* spin could be constrained by means of X-ray spectroscopic data. Future work will thus be devoted to fit the ion torus prediction to X-ray data:  this is the first way of constraining Sgr~A* spin.
 
The right panel of Fig.~\ref{fig:torus} shows two superimposed images of the ion torus at two different values of spin. This clearly demonstrates the fact already mentioned above that the torus shrinks at higher spin. What is mostly interesting in this Figure is that the diameter of the black hole silhouette (the thin circle of illuminated pixels at the center of the torus images) depends on the spin value: there is a difference of 3~$\mu$as in angular size for the silhouette of black holes with spin $0.5$ and $0.9$. This is very interesting as the angular size of the silhouette is nothing but the projection of the black hole's event horizon on the observer's sky. This size is thus independent on the other parameters describing the ion torus. If the angular size of the black hole silhouette can be constrained by observations, this would lead to a robust constraint on Sgr~A* spin. Such a constraint at the $\mu$as scale should be within reach of the future VLBI Event Horizon Telescope~\citep[see e.g.][]{broderick11}: this is the second way of constraining Sgr~A* spin.

 \section{Ray-tracing in the vicinity of non-Kerr compact objects}
 \label{3+1}
 
 One of the specificities of the GYOTO code is its ability to perform ray-tracing in non-analytic metrics, different from the Kerr case. Here, we give the first examples of images computed by GYOTO in such numerically computed metrics.
 
 When used with an analytical Kerr metric, GYOTO integrates the standard 4-dimensional equations of geodesics. When used with a numerically computed metric, GYOTO uses the 3+1 quantities describing the spacetime geometry, the 3-metric, lapse, shift and extrinsic curvature. The 4-dimensional geodesic equation can then be recast to a 3+1 equivalent that is integrated by GYOTO. For details, the reader is referred to~\citet{vincent12}.
 
\begin{figure}
\centering
	\includegraphics[width=0.225\textwidth]{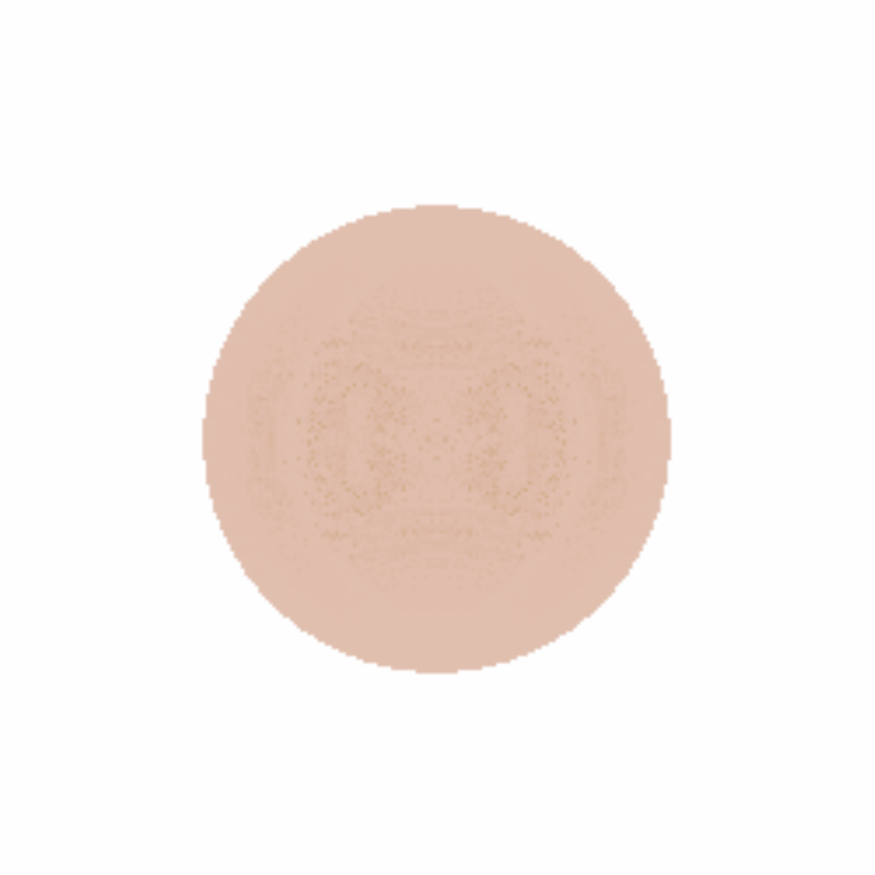}
	\includegraphics[width=0.225\textwidth]{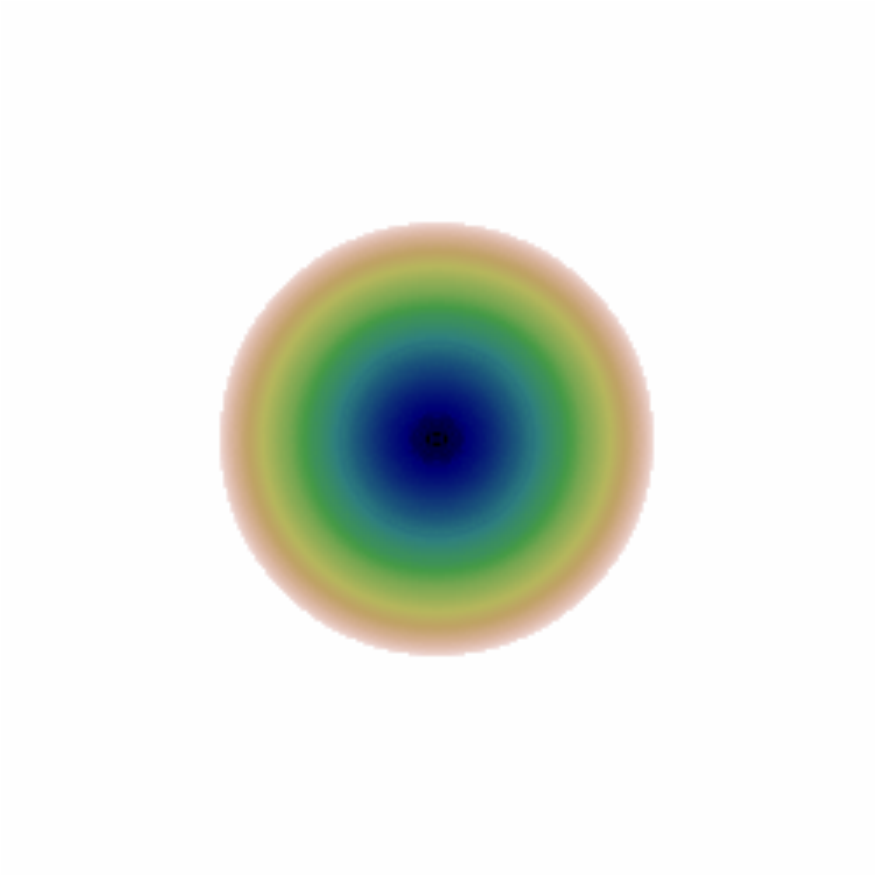}
	\includegraphics[width=0.225\textwidth]{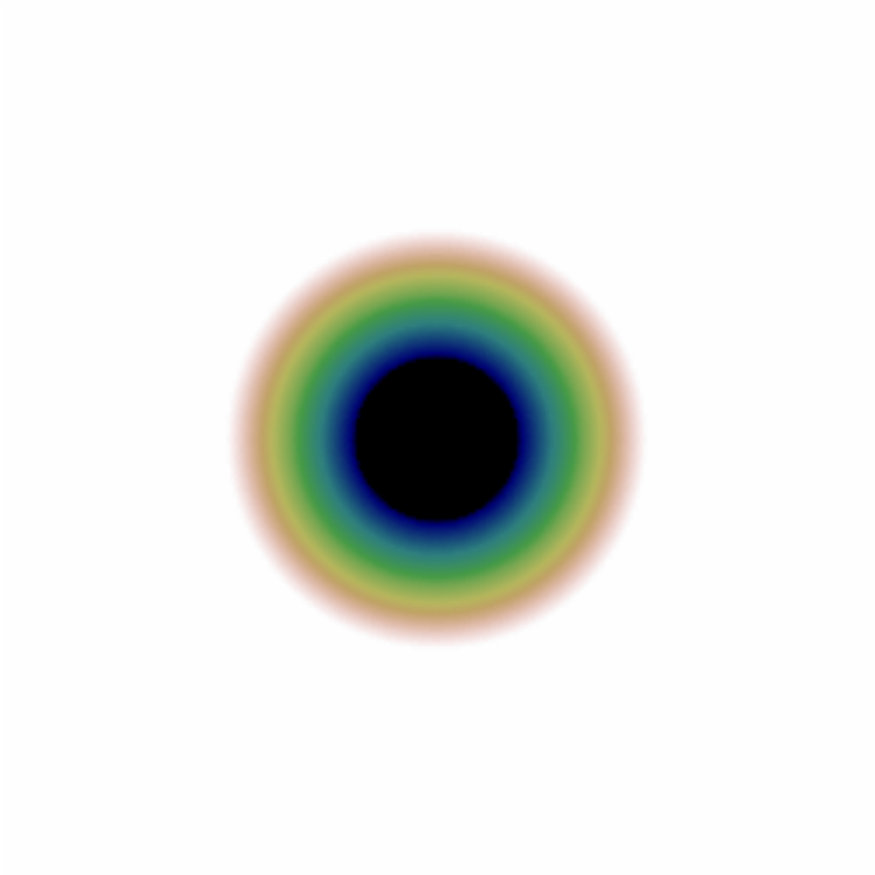}
	\includegraphics[width=0.225\textwidth]{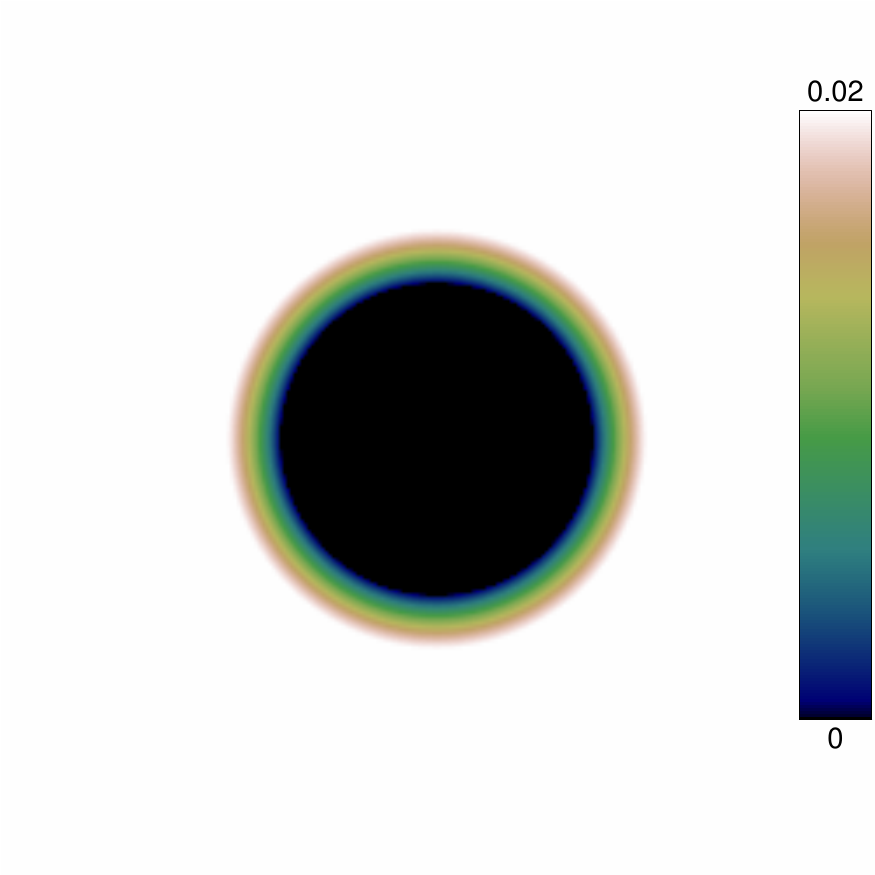}
	\caption{Images (i.e. map of specific intensity) of a non-rotating collapsing neutron star, with an optically thick surface emitting black body radiation at $10^{6}$~K. The color bar is common to the four panels and is given in SI units, $\mathrm{J\,s^{-1}\,m^{-2}\,ster^{-1}\,Hz^{-1}}$. The frequency of the photons in the observer's frame is chosen to be $10^{17}$~Hz, close to the maximum of the Planck function at $10^{6}$~K.}
	\label{f:dynaNS}
\end{figure}

Fig.~\ref{f:dynaNS} shows four successive images of a collapsing spherically symmetric neutron star, assumed to be optically thick and emitting as a blackbody with temperature $10^{6}$~K. The evolving metrics of the collapsing star are computed by means of the \texttt{CoCoNuT} code~\citep{dimmelmeier05}, and the ray-tracing is performed by GYOTO. This Figure shows the first example of images computed in numerical spacetimes, in a realistic (although simplified) astrophysical context. The growing of the event horizon is very clear on the different panels. It appears first at the center of the image as photons reaching the center of the observer's screen are emitted by closer parts of the star: they reach the observer sooner. It could seem strange that the shrinking of the neutron star is so tiny between the left panel (very beginning of the collapse) and the right panel (when nearly the whole star has disappeared behind the event horizon). However, this is due to the strong bending of null geodesics in the vicinity of the collapsing star that makes the apparent angular size of the image much bigger than the simple projection of the star (as if in flat spacetime). The same effect makes the angular size of a black hole shadow (as those depicted in the right panel of Fig.~\ref{fig:torus}) appear bigger than the projected size of the event horizon.

Fig.~\ref{f:dynaNS} is an illustration of what kind of computation GYOTO can do in numerical spacetimes. It is not (yet) relevant for astrophysical purposes. A natural development of such works would be to compute light curves and spectra of astrophysical phenomena in the vicinity of Sgr~A*, assuming this object is no longer a black hole of general relativity, but an alternative compact object. The numerical relativity group at Observatoire de Paris/LUTH is currently developing numerical metrics of such an alternative object: a boson star. It is a fascinating perspective to be able in the near future to develop simulations of observations of a probe phenomenon (such as a radiation flare) assuming the central object is either a Kerr black hole or a rotating boson star. Such simulations will allow to determine whether there is any observable differences between the two scenarii, and what kind of instrument could put to the light such a difference: this is the third way of constraining Sgr~A*'s nature.
  
\section{Conclusions}

Near-future instruments should allow to give reach to the immediate vicinity of black holes, that is to say, to the direct probing of strong-field general relativistic phenomena.

This paper presents three different ways of constraining the nature of the Galactic center black hole Sgr~A*, by means of current or near-future instruments. Assuming that Sgr~A* is surrounded by an ion torus, it shows that spectroscopic measurements in the X-rays could allow constraining the black hole spin. The same goal may be achieved by determining the angular size of Sgr~A* shadow by using the VLBI Event Horizon Telescope. Finally, the very nature of the central compact object may be probed by performing simulations of observations using alternative compact objects metrics.

\bibliographystyle{aa}  % A&A bibliography style file (aa.bst)
\bibliography{vincent} % your references in file: Yourfile.bib

\end{document}